\documentclass[a4paper,fleqn]{cas-dc}

\usepackage{amsmath}
\usepackage{amssymb}
\usepackage{bm}
\usepackage{dsfont}
\usepackage{natbib}
\usepackage{lineno}

\newcommand{\parens}[1]{\left(#1\right)}
\newcommand{\brackets}[1]{\left[#1\right]}

\newcommand{\chevrons}[1]{\left\langle#1\right\rangle}
\newcommand{\chevronsi}[1]{\langle#1\rangle}
\newcommand{\abs}[1]{\left|#1\right|}

\newcommand{\unit}[1]{\hat{\bm{#1}}}

\begin{document}
\let\WriteBookmarks\relax
\def\floatpagepagefraction{1}
\def\textpagefraction{.001}
\shorttitle{A Statistical Mechanical Ring Model}
\shortauthors{Jack T. Dinsmore}

\title [mode = title]{A Statistical Mechanics Model for Stable Rings Outside the Roche Limit}

\author[1]{Jack T. Dinsmore}[orcid=0000-0002-6401-778X]
\cormark[1]
\ead{jtd@stanford.edu}

\affiliation[1]{organization={Department of Physics and the Kavli Institute for Particle Astrophysics and Cosmology, Stanford University},
                city={Stanford},
                citysep={}, 
                postcode={94305}, 
                state={CA},
                country={USA}}

\begin{abstract}
Stellar occultations have revealed two rings around the dwarf planet Quaoar outside the Roche limit. Simulations suggest that the rings are sustained by large velocity dispersions. We present the first analytical, first-principles theory that extends the Roche limit to describe planetary rings with large velocity dispersion. The underlying principle is an analogy between a liquid/gas phase transition and the moon/ring transition: just as high temperatures can boil a liquid under pressure, high velocity dispersions can stabilize ring systems beyond the Roche limit. We employ statistical mechanics to derive a modification to the Roche limit that treats rings with nonzero velocity dispersion. We demonstrate the new model's consistency with previous simulations of Quaoar's outermost ring. We also note that dense regions of the ring experience negative pressure, which would cause them to shrink over time if the assumptions of the model continue to hold. This could help explain the rings' azimuthal asymmetry.
\end{abstract}


\begin{highlights}
\item Finds a first-principles extension to the Roche-limit for high velocity dispersion rings
\item Shows that the predictions agree with simulations of Quaoar's outermost ring
\item Introduces a new statistical mechanics theoretical framework for planetary rings
\item Gives a possible alternate explanation for Q1R's azimuthal asymmetry
\end{highlights}


\maketitle

\section{Introduction} \label{sec:intro}
The discovery of narrow rings around the Centaur Chariklo \citep{braga2014ring}, the dwarf planets Haumea  \citep{ortiz2017size} and Quaoar \citep{morgado2023dense,pereira2023two}, and possibly the Centaur Chiron \citep{ruprecht2015november, sickafoose2020characterization, ortiz2023changing, Pereira2025rings} via stellar occultations have sparked much curiosity. The rings appear radially confined, possibly by the existence of undiscovered shepherd satellites \citep{braga2014ring,michikoshi2017simulating,sickafoose2024numerical}, or spin orbit resonances (SORs) with the central body \citep{sicardy2019ring, morgado2023dense, salo2026rings}. In the case of Quaoar's outermost ring Q1R, the 6/1 mean motion resonance (MMR) with Quaoar's moon Weywot \citep{rodriguez2023dynamical, morgado2023dense} and 5/3 MMR with a newly discovered moon \citep{proudfoot2025orbital} may also contribute to radial confinement. Chariklo's and Quaoar's rings show azimuthal asymmetry which could also be driven by resonances \citep{morgado2023dense}, though long-lived asymmetry can also be created by apse-alignment caused by the rings' self-gravity \citep{goldreich1979precession, goldreich1979towards, pan2016on,melita2020apse}.

Haumea's and Chariklo's rings are inside the classical Roche limit, where tidal forces are strong enough to prevent the rings from collapsing into a satellite under inter-particle gravitational forces. However, Quaoar's two rings Q1R and Q2R lie outside the Roche limit for reasonable particle densities---the first dense rings discovered to do so.

\citet{morgado2023dense} suggest that these rings are supported by large velocity dispersion $\sigma$, and are stable when $\sigma \gtrsim$ the two-body escape velocity. Furthermore, velocity dispersion can resist gravitational collapse for a self-gravitating gaseous disk if the Toomre-$Q$ parameter satisfies $Q>1$ \citep{toomre1964on,goldreich1965spiral}. While both of these criteria predict stability at large velocity dispersion, they do not reproduce the Roche limit at small velocity dispersion. Thus they cannot be viewed as extensions to the Roche limit for high velocity dispersion rings.

This work builds on these models to extend the Roche limit to rings with high velocity dispersion. \S\ref{sec:dispersion} modifies the Toomre-$Q$ analysis to derive a simple stability criterion which is both consistent with the Roche limit at low $\sigma$ and pushes the limit out at high $\sigma$. However, this modification does not provide a good fit to simulations. We therefore propose a statistical mechanics method to make the final predictions. \S\ref{sec:statistics} describes the statistical model and its assumptions, arguing that it is more accurate than the modified Toomre-$Q$ analysis and can treat variety of problems beyond the stability criterion addressed by this work. The model's full details are given in \S\ref{sec:model}, and some illustrative calculations are presented in \S\ref{sec:results}. \S\ref{sec:roche} computes the Roche limit modification taking velocity dispersion into account, and comparison with ring simulations indicate that the formula is accurate. \S\ref{sec:pressure} then applies the model to dense portions of velocity-supported rings and finds evidence that they may shrink over time. The paper concludes with \S\ref{sec:conclusion}.

\section{Justification of the Statistical Model}
We define the ratio between the ring radius $a$ and the classical Roche limit $a_\mathrm{Roche}$ to be
\begin{equation}
  \zeta = \frac{a}{a_\mathrm{Roche}} = \frac{a}{R_M}\parens{\frac{\rho_m}{12\rho_M}}^{1/3},
  \label{eqn:zeta}
\end{equation}
where $R_M$ is the central body radius and $\rho_m$ and $\rho_M$ are the densities of the ring particles and central body. We have set $a_\mathrm{Roche}$ equal to the ring radius at which the mutual Hill radius of two abutting particles is equal to their center-of-mass separation, following \citet{morgado2023dense}. 

\subsection{Dispersion Relation Approach}
\label{sec:dispersion}

This section presents a simple way to modify the Roche limit to explain high velocity dispersion rings using the dispersion relation techniques of the Toomre-$Q$ criterion, though we will show that the result is not as accurate as the statistical model discussed in the next section. 
Consider a ring containing spherical particles of mass $m$ and radius $R$ at optical depth $\tau$ and orbital frequency $\Omega$. We begin by approximating the ring as a self-gravitating, Keplerian fluid of surface density $\Sigma = m\tau / (\pi R^2)$ with sound speed equal to the ring's radial velocity dispersion $\sigma$. Waves with frequency $\omega$ and wavenumber $\bm k$ propagate with dispersion relation $\omega^2 = \Omega^2 - 2\pi G\Sigma |\bm k| + \sigma^2 |\bm k|^2$ \citep{binney2008galactic}. Modes where the right hand side is negative are unstable, which gives the the stability criterion for an individual mode:
\begin{equation}
  \zeta < \frac{1}{\tau^{1/3}}\brackets{\frac{1}{24R|\bm k|} + \frac{\sigma^2}{R^2 \Omega^2} \frac{R|\bm k|}{24}}^{1/3}.
  \label{eqn:stability-all-k}
\end{equation}

The Toomre-$Q$ criterion assumes that the system is globally stable only when all modes are stable \citep[e.g.][]{toomre1964on,goldreich1965spiral}, but for rings, gravitational collapse starts at small scales where individual particles accrete. Suppose that ring stability is controlled by only one, small-scale, mode. Then Eq.~\ref{eqn:stability-all-k} is the ring's global stability criterion after $|\bm k|$ is replaced by that mode's wavenumber---say $k_\mathrm{moon}$. Setting $k_\mathrm{moon}=1/(6 R)$ would lead to the stability criterion
\begin{equation}
\zeta < \frac{1}{(4\tau)^{1/3}}\brackets{1 + \frac{1}{36}\frac{\sigma^2}{R^2 \Omega^2}}^{1/3}.
\label{eqn:stability-disp}
\end{equation}
which reproduces the classical Roche limit $\zeta < 1$ at $\sigma = 0$ (for $\tau = 1/4$). This criterion also predicts stable rings beyond the Roche limit given high $\sigma$, demonstrating that stable \textit{small-scale} density perturbations may explain Quaoar's rings' stability. However, Eq.~\ref{eqn:stability-disp} does not reproduce the Roche limit when $\tau \neq 1/4$ due to the dependence on optical depth, and it does not reproduce simulations even when $\tau = 1/4$ as discussed next.

\begin{figure}
  \centering
  \includegraphics[width=\linewidth]{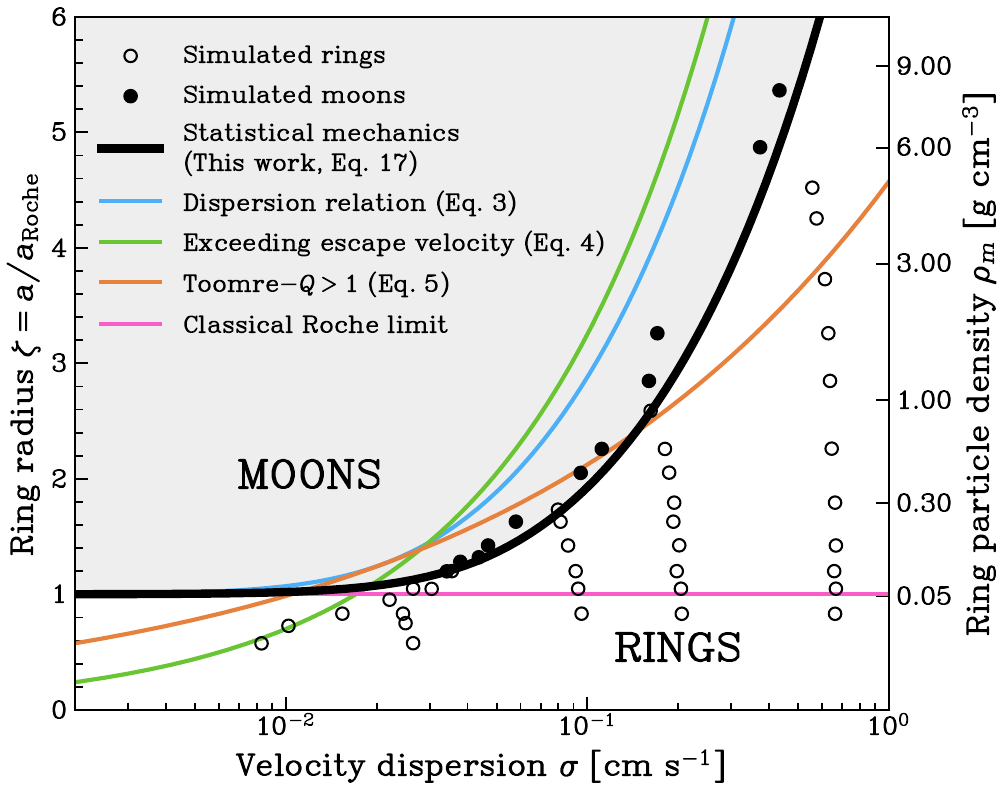}
  \caption{
    Different stability criteria for planetary rings as a function of velocity dispersion, compared to simulations of Q1R \citep{morgado2023dense}. The open points represent simulated rings while the closed points collapsed to moons. The $\zeta$ values of the ring are converted to physical particle densities (right axis) using the properties of the Quaoar system. The simplest models (green, pink, and orange curves) have been previously suggested, while the blue curve is novel to this analysis. None of these reproduce the simulation results. The statistical mechanics presented by this work (black curve) is the only model to fit the simulations.
  }
  \label{fig:models}
\end{figure}

Fig.~\ref{fig:models} presents all the previously mentioned stability criteria  as colored lines. This includes Eq.~\ref{eqn:stability-disp}, the classical Roche limit, and the $\sigma > v_\mathrm{esc}$ and Toomre-$Q > 1$ criteria mentioned in \S\ref{sec:intro}. Converted to conditions on $\zeta$, the latter two are
\begin{align}
  \zeta &< \parens{\frac{\sigma^2}{24\Omega^2 R^2}}^{1/3} \qquad (\sigma > v_\mathrm{esc})\label{eqn:esc},\\
  \zeta &< \parens{\frac{\sigma}{12\Omega R \tau}}^{1/3} \qquad (Q > 1) \label{eqn:toomre}
\end{align}
where we have used the two-body escape velocity $v_\mathrm{esc} = \sqrt{2Gm/R}$. 
\citet{morgado2023dense} conducted simulations of Q1R at different particle densities and coefficients of restitution, which exhibited different velocity dispersions in steady state. We convert the particle density to $\zeta$ using $M=1.212\times 10^{21}$ kg \citep{proudfoot2025orbital}, $a=4057$ km \citep{pereira2023two}, and the simulations' particle radius of $R = 1$ m. Whether these simulations resulted in stable rings/moons are shown as open/closed circles. All curves assume the simulation value of $\tau=1/4$. None of these models match the simulations: the $\sigma > v_\mathrm{esc}$ and $Q > 1$ models do not reproduce the Roche limit at small $\sigma$, and Eq.~\ref{eqn:stability-disp} fails to match the high-$\sigma$ points. We suggest that the the $\sigma > v_\mathrm{esc}$ and $Q>1$ curves fail because they do not model the scale of particle accretion which critically determines the Roche limit, while the initial assumption that the ring is a fluid drives the inaccuracy of Eq.~\ref{eqn:stability-disp}. We are therefore motivated to find another formalism for extending the Roche limit which considers both effects. This formalism will yield the black curve, which does fit the simulated points.

\subsection{Statistical Mechanics Approach}
\label{sec:statistics}
We motivate a statistical model with an analogy between gravitational collapse in planetary rings and a gas-liquid phase transition. For sufficiently simple systems, one can compute the temperature at which a gas condenses into a liquid due to inter-molecular forces without simulating the detailed, dynamical interactions between particles. This is enabled by the simplifying assumptions of equilibrium and ergodicity in statistical mechanics (discussed further below). Using the same techniques, the statistical method will compute the ``effective temperature'' (or velocity dispersion) at which a ring collapses into a moon under gravitational forces.

The statistical model we introduce considers the effects of epicyclic motion, self-gravity, and velocity dispersion as did the dispersion relation method in the previous section, while achieving other advantages. This statistical model straightforwardly treats finite particle size (a weakness of the fluid picture in \S\ref{sec:dispersion}). Additionally, the model could be used to compute from first principles the ring's pressure, the response to perturbing potentials, and the effects of large density perturbations, which are not accessible to dispersion relation approaches. These tools may prove useful for future studies seeking to explore the other unusual features of dwarf planet rings.

The assumptions of this model are as follows.
\begin{enumerate}
  \item \textit{Equilibrium}: We assume detailed balance, or that a process and its reverse process occur at equal rates.
  \item \textit{Small density perturbations}: Given a surface density perturbation of size $\delta \Sigma$, the current model considers effects of order $\delta \Sigma$ but not $\delta \Sigma^2$ or higher.
  \item \textit{Ergodicity}: We assume that after enough time, the particles eventually achieve all possible eccentricities, inclinations, and arguments of the pericenter and ascending node.
\end{enumerate}
The assumption of equilibrium is necessary for this statistical model, while assumption 2 has been added for simplicity. It is also necessary to assume ergodicity for some parameters, and our reasoning behind this specific choice of parameters is given in \S\ref{sec:dof}.

These assumptions are all violated in some scenarios, as described below. When the violations are important, kinetic approaches should be used instead of statistical models.

Inelastic collisions routinely dissipate energy while viscous mixing or resonances inject it, and balancing the total energy fluxes will lead to a steady-state velocity dispersion. Our statistical model captures this important effect via the ``effective temperature.'' But these energy fluxes in fact vary over the ring, depending on local particle density. This variation violates assumption 1 and will perturb the ring properties away from the values predicted by statistical mechanics, which we do not model. However, gravitational interactions do satisfy detailed balance and will drive the ring properties back to their statistically expected values. The assumption of equilibrium should be valid when gravitational interactions dominate collisions, which occurs when optical depth is low or the coefficient of restitution is near unity.

Swing amplification \citep{julian1966non,salo1992gravitational} shapes density perturbations in rings. Since this effect involves the influence of a density perturbation's shape on itself, it is second order in density and will not appear in this first-order approach under assumption 2. Our predictions for the shape of strong over-densities will therefore be inaccurate. Second order effects could be added with a diagrammatic expansion as has been done for other statistical mechanical systems, albeit with substantially more computational effort.

Due to assumption 3, the predictions of the statistical model will be inaccurate over short time scales, when the system does not have time to explore all possible states. Simulations of Chariklo have shown that ring properties can vary rapidly \citep[e.g.][]{michikoshi2017simulating}, which this model cannot describe.

In summary the model does not correctly describe optically thick regions, strong over-densities, and time-variable phenomena. However, the statistical model presents a simple and powerful tool when studying time-independent phenomena, for smooth and optically thin rings.

\section{Details of the Statistical Model} \label{sec:model}
We employ the canonical ensemble, which fundamentally supposes that the probability for the ring to occupy some state $\mathcal{S}$ is proportional to the Boltzmann factor $e^{-H(\mathcal{S})/T}$ \citep{huang1987statistical}. $H(\mathcal{S})$ is the state's energy, and $T$ is an effective temperature which encodes velocity dispersion (discussed further in \S\ref{sec:velocity}).\footnote{We set Boltzmann's constant to one, so $T$ has units of energy.} Using this weight assumes equilibrium. Then invoking ergodicity, the time-average of an arbitrary ring property $Q$ can then be computed with a statistical average over the ring states, weighted by this factor:
\begin{equation}
  \chevronsi{Q} = \frac{\sum_\mathcal{S} Q(\mathcal{S}) e^{-H(\mathcal{S})/T}}{\sum_\mathcal{S} e^{-H(\mathcal{S})/T}}.
  \label{eqn:average}
\end{equation}

\subsection{Parametrization of the States} \label{sec:dof}
How the states are parametrized is crucial because the final results implicitly assume ergodicity and detailed balance for the chosen states. For example, the simple choice of defining a state by the positions and velocities of all particles would yield poor results because epicyclic motion causes these values to evolve over time, violating detailed balance. The particles' orbital parameters are constant up to gravitational or collisional perturbations, so using them avoids the problem.

However, if we treated the particles as able to achieve any semi-major axis, the states with particle semi-major axes $a_j=0$ (where $j$ indexes over particles) would dominate the sum for $\chevronsi{Q}$ because their orbital energy is $-\infty$. In reality, the ring particles have a narrow range of semi-major axes that changes only very slowly as the ring radially expands. This is because each particle's $z$-component\footnote{We use Hill coordinates, defined such that the origin co-rotates with the middle of the ring at semi-major axis $a$, with $\unit x$ pointing away from the central body, $\unit y$ forward in the orbit, and $\unit z$ normal to the orbit.} of orbital angular momentum $L_{z,j}$ is conserved up to gravitational interactions and collisions, which prevents $L_{z,j}$ from reaching the high values necessary to attain these lowest-energy states. We model this by assuming that the particles' orbital guiding centers $\bm s_j$ are uniformly distributed within the ring, where $\bm s_j$ is the point following a circular orbit in the ring plane with orbital angular momentum equal to $L_{z,j}\unit z$. We assume the remaining orbital elements (inclination $\iota_j$, eccentricity $e_j$, and longitudes of the ascending node $\Omega_j$ and pericenter $\varpi_j$) are distributed by the Boltzmann factor.

For later simplicity, we define the complex variables $c_j = e_j\exp[i\varpi_j]$ and $h_j = \iota_j \exp[i \Omega_j]$. (Note the distinction between the inclination $\iota_j$ and the imaginary unit $i$.) The particle's displacement from its guiding center at mean anomaly $M_j$ is
\begin{equation}
  \begin{aligned}
    \bm r_j =a \big[&-\unit x \mathrm{Re}(c_j\exp[iM_j]) + 2\unit y \mathrm{Im}(c_j\exp[iM_j])\\
    &+\unit z \mathrm{Im}(h_j\exp[iM_j])\big]
  \end{aligned}
  \label{eqn:displacement}
\end{equation}
to first order in $c_j$ and $h_j$. We then define the fields $c(\bm x)$ and $h(\bm x)$ where $\bm x$ is a position on the ring's orbital plane, such that $c_j = c(\bm s_j)$. These fields form the ring states $\mathcal{S}$ appearing in Eq.~\ref{eqn:average}.\footnote{In order for averages over $c$ and $h$ to have meaning, Liouville's theorem must apply to them. While these variables are not canonical, their Poisson brackets are constant. They are therefore proportional to canonical variables for which Liouville's theorem does hold, which implies the theorem also applies to $c$ and $h$.}

\subsection{Computing the Average Values of $c$ and $h$}
The sums in Eq.~\ref{eqn:average} should avoid states $\mathcal{S}$ where particles overlap. This can be done by adding a penalty term to $H(\mathcal{S})$ to yield the ``effective energy'' $H_\mathrm{eff}(c, h)$ (derived in appendix \ref{app:heff}), which depends on the values of the $c$ and $h$ fields. With the penalty term, Eq.~\ref{eqn:average} becomes
\begin{equation}
  \chevronsi{Q} = \frac{\sum_c\sum_h Q(c,h) e^{-H_\mathrm{eff}(c, h)/T}}{\sum_c\sum_h e^{-H_\mathrm{eff}(c, h)/T}}
  \label{eqn:average2}
\end{equation}
where the sums are now over all configurations of the fields. This formula can now be explicitly evaluated given $H_\mathrm{eff}(c, h)$. Neglecting high order terms in eccentricity or inclination (assumption 2), $H_\mathrm{eff}(c, h)$ can be expressed in the form
\begin{equation}
  H_\mathrm{eff}(c,h) = \int_{|\bm k| < \Lambda} \frac{d^2\bm k}{(2\pi)^2} \brackets{\epsilon_c(\bm k) |c(\bm k)|^2 + \epsilon_h(\bm k) |h(\bm k)|^2} + C
  \label{eqn:h}
\end{equation}
where $c(\bm k)$ and $h(\bm k)$ are the Fourier transforms\footnote{We use the Fourier transform convention $c(\bm k) = \int d^2 \bm x\, e^{-i\bm k \cdot \bm x} c(\bm x)$, with inverse $c(\bm x) = \int \frac{d^2 \bm k}{(2\pi)^2}\, e^{i\bm k \cdot \bm x} c(\bm k)$.} of the $c$ and $h$ fields. The functions $\epsilon_c(\bm k)$ and $\epsilon_h(\bm k)$ are the energies of those modes per unit amplitude. $H_\mathrm{eff}$ resembles the energy of a harmonic oscillator, so henceforth we call $\epsilon_{c/h}$ the ``effective spring constants'' for modes of wavenumber $\bm k$. The constant energy offset $C$ in Eq.~\ref{eqn:h} does not come into predictions of ring properties because constants factor out of Eq.~\ref{eqn:average2}.

The area element $d^2\bm{k}$ integrates over all wavenumbers $\bm k$. We have inserted a large-$k$ cutoff $\Lambda$ because field fluctuations much smaller than particles' individual size do not affect the values of the particles' $c_j$ and $h_j$. For example, if the particles were arranged in a square grid, the inter-particle spacing would be $\ell = R \sqrt{\pi / \tau}$ and the cutoff would be the Nyquist limit $\Lambda = \pi / \ell = \sqrt{\pi\tau} / R$. We use this Nyquist value, but a simulation or more advanced theoretical tools would be required to handle the correct high-$k$ behavior.

Appendix \ref{app:h} computes the ring energy, determining the effective spring constants to be
\begin{align}
  &\begin{aligned}
    \epsilon_c(\bm k) = \frac{ma^2\Omega^2\tau}{2\pi R^2}\bigg[1 + &\parens{\frac{T}{2 m\Omega^2 R^2}\frac{1}{1-\tau} - \frac{12\tau\zeta^3}{R|\bm k|}}\\
    &\times R^2(k_x^2 + 4k_y^2)\mathbb{1}(\bm k)\bigg],
  \end{aligned} \label{eqn:eps-c}\\
  &\epsilon_h(\bm k) = \frac{ma^2\Omega^2 \tau}{2\pi R^2}\brackets{1 +12\tau\zeta^3 R|\bm k|}. \label{eqn:eps-h}
\end{align}
The first term in both equations denotes the orbital energy of each particle. The terms proportional to $\zeta^3 \tau$ describe the gravitational potential energy between particles. The previously mentioned penalty term, which avoids counting states where particles collide, is the term in Eq.~\ref{eqn:eps-c} proportional to $T$. (The function $\mathbb{1}(\bm k)$ is equal to 1, except in steady state for $\bm k$ such that $k_x^2 + 4k_y^2 > m\Omega^2 /T$, in which case $\mathbb{1}(\bm k)=0$. See appendix \ref{app:h} for the derivation.)

While Eq.~\ref{eqn:average2} can be used to compute $\chevronsi{Q}$ for any ring property $Q$, the conclusions of this work require only the mean-squared amplitudes of eccentricity and inclination fluctuations $\chevronsi{|c(\bm k)|^2}$ and $\chevronsi{|h(\bm k)|^2}$. The sum of Eq.~\ref{eqn:average2} can then be avoided in favor of the equipartition theorem. For quadratic energies such as Eq.~\ref{eqn:h}, the equipartition theorem dictates that the average energy stored in each degree of freedom is $AT/2$, where $A$ is the area of the ring \citep{huang1987statistical}. Since $|c(\bm k)|^2$ and $|h(\bm k)|^2$ contain two degrees of freedom each from the real and complex parts,
\begin{equation}
  \chevrons{|c(\bm k)|^2} = \frac{TA}{\epsilon_c(\bm k)}, \qquad\chevrons{|h(\bm k)|^2} = \frac{TA}{\epsilon_h(\bm k)}.
  \label{eqn:exp}
\end{equation}
The modes with larger effective spring constants $\epsilon_{c/h}$ achieve smaller amplitudes on average.

\section{Results} \label{sec:results}

\subsection{Velocity Dispersion}\label{sec:velocity}
One can write particle velocity $\bm v_j = \Omega \partial \bm r_j / \partial M_j$ in terms of the particle's $c_j$ and $h_j$ variables, and mean anomaly $M_j$ using Eq.~\ref{eqn:displacement}, and therefore obtain the mean-square velocity $\chevronsi{v^2}$:
\begin{equation}
  \begin{aligned}
    \chevrons{v^2} &= \frac{\Omega^2a^2}{A}\int_{|\bm k| < \Lambda} \frac{d^2\bm k}{(2\pi)^2} \brackets{\frac{5}{2}\chevrons{|c(\bm k)|^2}+\frac{1}{2}\chevrons{|h(\bm k)|^2}}\\
    &= \frac{\Omega^2a^2T}{2}\int_{|\bm k| < \Lambda} \frac{d^2\bm k}{(2\pi)^2} \brackets{\frac{5}{\epsilon_c(\bm k)}+\frac{1}{\epsilon_h(\bm k)}}\\
  \end{aligned}
  \label{eqn:v}
\end{equation}
where the second line follows from Eq.~\ref{eqn:exp}. The mean-square relative velocity between particles is $2\chevronsi{v^2}$, the radial component of which is $(2/3)\chevronsi{v^2}$ assuming isotropic velocities. Eq.~\ref{eqn:v} cannot be integrated exactly due to the $k$-dependent terms in $\epsilon_c$ and $\epsilon_h$, but these can be ignored to a good approximation, leaving
\begin{equation}
  \sigma \approx \parens{\frac{2}{3}\chevrons{v^2}}^{1/2} \approx \sqrt{\frac{\pi T}{m}}.
  \label{eqn:sigma}
\end{equation}
This equation gives the connection between velocity dispersion $\sigma$ and effective temperature $T$.

\subsection{Density Correlations} \label{sec:correlation}

\begin{figure}
  \centering
  \includegraphics[width=\linewidth]{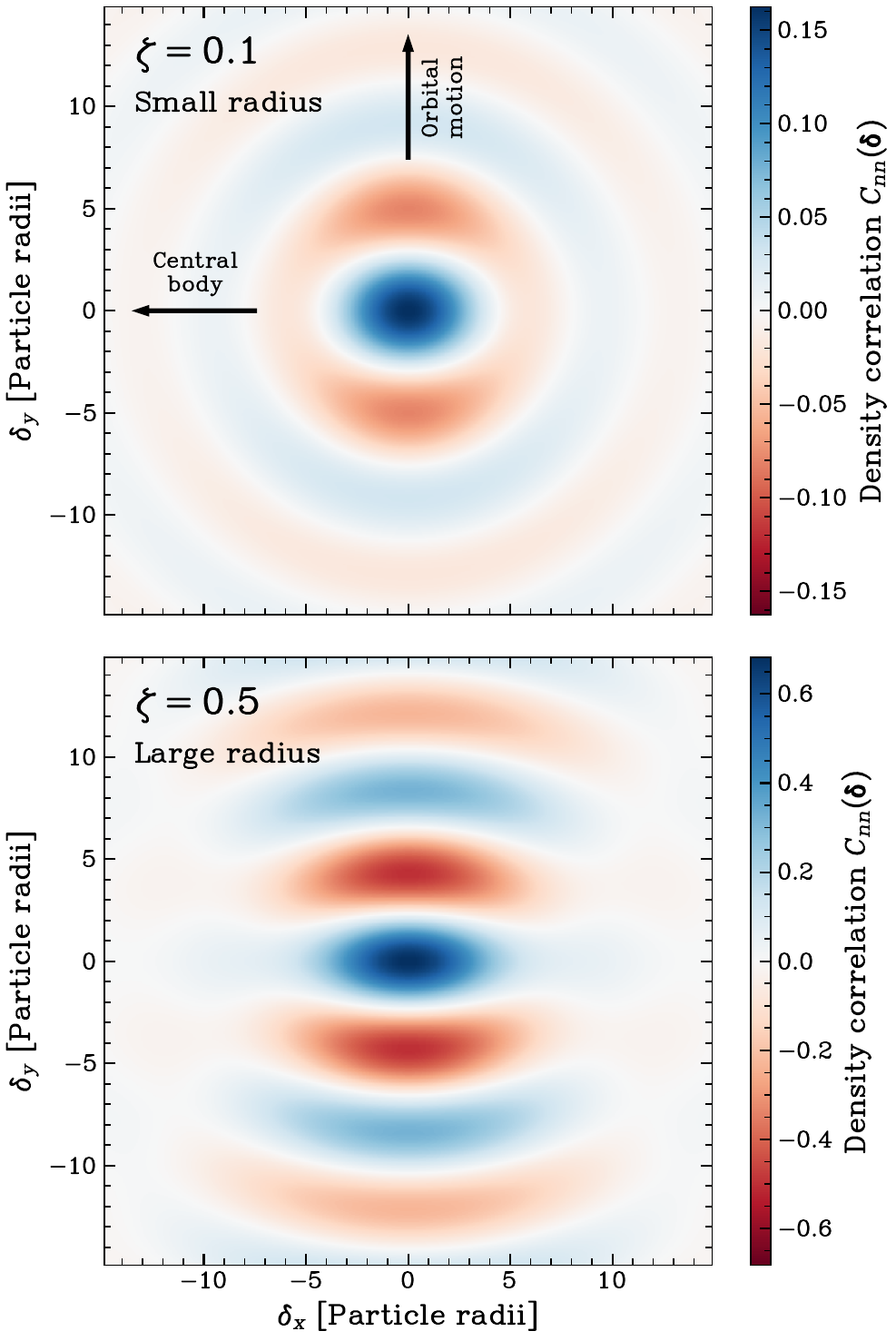}
  \caption{Gravity-induced density correlations in the orbital plane, viewed from above the orbital plane. Values are shown for a ring well inside/just inside the Roche limit of $\zeta = 1$ (top/bottom). The stronger correlations in the bottom image indicate density inhomogeneity in equilibrium. The over-dense structures are elongated along the $x$-axis nearer to the Roche limit, but with the current simplified model, the full shape of density wakes is not reproduced.}
  \label{fig:density}
\end{figure}
One application of this model is to predict the shape of local over-densities in the ring, which we can use to probe the validity of the assumptions. The self-gravity-induced density correlation between two points separated by distance $\bm \delta$ is
\begin{equation}
  \begin{aligned}
    C_{nn}(\bm \delta) &= \frac{1}{A}\int d^2\bm x\, \chevrons{\frac{\delta n(\bm x)\delta n(\bm x+\bm \delta)}{n_0^2}} \\
    &= \frac{1}{A}\int \frac{d^2\bm k}{(2\pi)^2}\, \chevrons{\abs{\frac{\delta n(\bm k)}{n_0}}^2} e^{i\bm k\cdot \bm \delta},
  \end{aligned}
  \label{eqn:corr}
\end{equation}
where $\delta n(\bm x)$ is the perturbation to the number density of particles at position $\bm x$. Eq.~\ref{eqn:n} relates $\delta n(\bm k)$ to the eccentricity field $c$, so $C_{nn}$ can be numerically integrated using $\chevronsi{|c(\bm k)|^2}$ given by Eq.~\ref{eqn:exp}. Fig.~\ref{fig:density} presents example density correlations for $T=mR^2 \Omega^2/4$ and $\tau = 1/4$. 

Near the central body (low $\zeta$, top panel), density correlations are quite low, suggesting that rings are largely uniform. Farther from the planet (bottom panel), the model shows stronger density correlations, with narrow structures elongated towards the central body. These are reminiscent of the density wakes seen in simulations of Saturn's \citep{salo1992gravitational} and Chariklo's \citep{michikoshi2017simulating, salo2026rings} rings. However, the simulated density wakes are larger and notably sheared whereas ours are not, likely because swing amplification and other effects of order $\delta \Sigma^2$ are not included in this model, nor are dynamical effects which may be necessary to correctly describe these over-densities. Thus, Fig.~\ref{fig:density} illustrates the limitations of the present model when describing significantly over-dense portions of the ring.

\begin{figure*}
  \centering
  \includegraphics[width=0.49\linewidth]{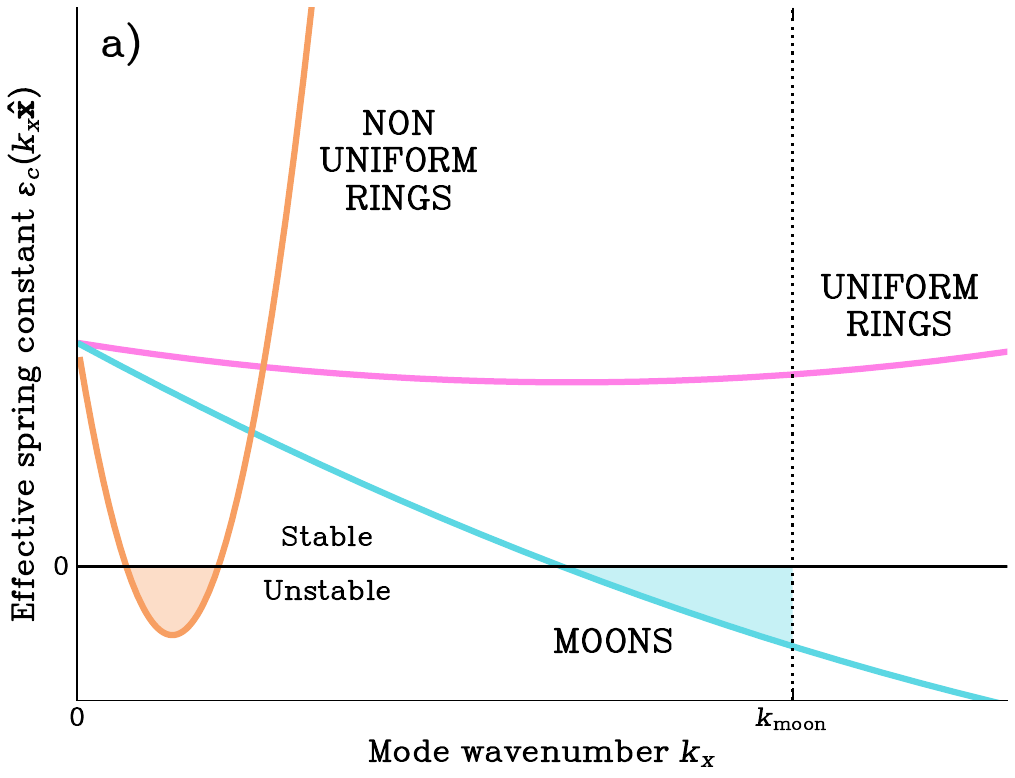}
  \includegraphics[width=0.49\linewidth]{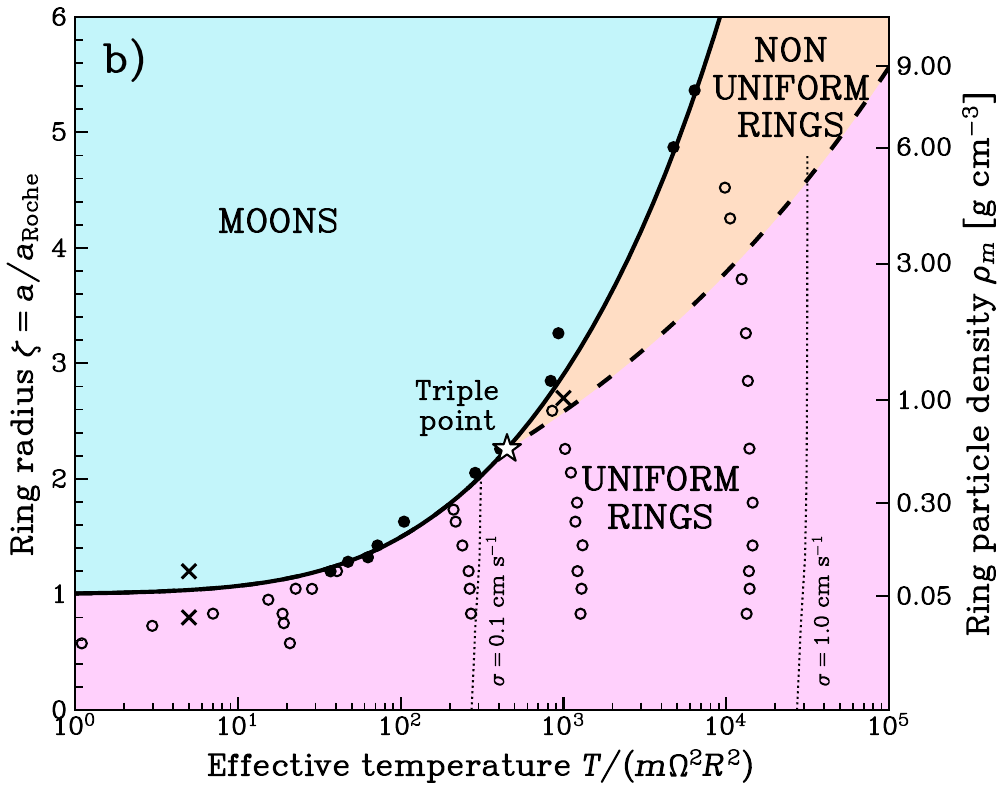}
  \caption{\textbf{(a)}: The effective spring constant $\epsilon_c(\bm k)$ (Eq.~\ref{eqn:eps-c}) as a function of wavenumber $\bm k$. The shaded regions have negative spring constant and are unstable. The shape of this curve dictates which phase the system occupies. An example is given for each phase, and the exact choices of effective temperature $T$ and $\zeta$ are marked on panel b with $\times$'s.
  \textbf{(b)}: Phase diagram of a planetary ring. The occupied phase is shown as a function of $T$ and $\zeta$. Dotted vertical lines show the corresponding velocity dispersion for the parameters of Q1R, and the right axis shows the particle density. Solid/hollow points represent simulations of Q1R made by \citet{morgado2023dense} which led to moons/rings. The black solid line shows the predicted boundary of the moon phase (Eq.~\ref{eqn:phase}), which matches simulations well. The simulations in the ``non-uniform'' rings (defined by Eq.~\ref{eqn:a-yellow} and delimited by the dashed line) appear stable, despite the presence of unstable modes as illustrated in panel a. The densest parts of Q1R could occupy this non-uniform ring state, which may have different long-term evolution than the uniform phase (see \S\ref{sec:pressure}).}
  \label{fig:phase}
\end{figure*}

Eq.~\ref{eqn:corr} predicts that density non-uniformities are smoothed out at high velocity dispersion since $\epsilon_c(\bm k)$ is large at these high temperatures, making $|c(\bm k)|^2$ and therefore $|\delta n(\bm k)|^2$ small. One therefore expects high velocity dispersion rings to display less severe density anisotropy, even when close to the outermost stable radius. The simulations of Q1R \citep{morgado2023dense} indeed depict less severe density anisotropy at higher velocity dispersion.

\section{Roche Limit Modification for High Velocity Dispersion} 
\label{sec:roche}
When the effective spring constant of eccentricity fluctuations $\epsilon_c(\bm k)$ (Eq.~\ref{eqn:eps-c}) is negative for some mode $\bm k$, increasing the power of particle eccentricities for this mode $|c(\bm k)|^2$ decreases the ring's total energy $H_\mathrm{eff}$. Therefore, it is energetically favorable for $|c(\bm k)|^2$ to increase exponentially until the ring collapses. However Keplerian shearing distorts this unstable mode as it grows. If the wavefronts are not tangent to the orbit, then the inner ends of the wavefronts are pulled forward in the orbit, and the outer ends are dragged behind. This wavefront shearing manifests in Fourier space as the wavenumber shifting at a rate $\bm {\dot k} = \frac{3}{2}\Omega k_y \unit x$, which could take the unstable $\bm k$ into a stable region of parameter space again before the ring collapses, halting the unstable growth.

In order to evaluate the ring's stability, we consider only the stability of modes with wavefronts tangent to the orbit ($k_y=0$), since shearing cannot shift them and cannot halt unstable growth. If $\epsilon_c(\bm k) < 0$, we therefore say these $k_y=0$ modes are unstable. Certainly other modes occur and could be unstable, which a more complicated analysis could take into account.

As in \S\ref{sec:dispersion}, we assume that the ring's stability depends purely on the stability at a specific wavenumber $k_\mathrm{moon}$, which denotes the scale at which moon particles begin to accrete. We define the stable systems ($\epsilon_c(k_\mathrm{moon}\unit x)>0$) to be in the ``ring'' phase, and the unstable systems ($\epsilon_c(k_\mathrm{moon} \unit x)<0$) to be in the ``moon'' phase.  We further subdivide the ring phase into two classes: systems where all modes with $k_x < k_\mathrm{moon}$ are stable occupy the ``uniform ring'' phase, and systems with some unstable modes in that range occupy ``non-uniform ring'' phase (discussed further below).

Fig.~\ref{fig:phase}a illustrates these definitions by showing $\epsilon_c(k_x \unit x)$ as a function of $k_x$ in each phase. The occupied phase is depicted in Fig.~\ref{fig:phase}b as a function of $\zeta$ and $T$. As in Fig.~\ref{fig:models}, solid/hollow points represent simulations of Q1R performed by \citet{morgado2023dense} which led to moons/rings.\footnote{Fig.~\ref{fig:phase}b uses the full $k$-dependent behavior of $\epsilon_c$ at the simulations' optical depth of $\tau=1/4$ to translate $\sigma$ into $T$. The resulting $\sigma$ contours are nearly independent of $\zeta$, indicating that the approximation used to simply relate $\sigma$ to $T$ via Eq.~\ref{eqn:sigma} (used elsewhere in this work) is fairly accurate.}

\paragraph{Uniform ring phase} Using Eq.~\ref{eqn:eps-c}, the ring phase's stability criterion can be rewritten in terms of $\zeta$ and $T$ to be
\begin{equation}
  \zeta < \parens{1 + \frac{T}{288 m\Omega^2 R^2 \tau^2 (1-\tau)}}^{1/3},
  \label{eqn:phase-1}
\end{equation}
where we have set $k_\mathrm{moon}$ to reproduce the classical Roche limit of $\zeta<1$ at $\sigma = 0$, as in \S\ref{sec:dispersion}. This inequality depends on optical depth $\tau$. Since $\tau$ fluctuates statistically, true stability should require that Eq.~\ref{eqn:phase-1} hold for all $\tau$. Note that the denominator of Eq.~\ref{eqn:phase-1} is maximized at $\tau = 2/3$. Substituting this value of $\tau$ gives the true, $\tau$-independent stability criterion---if this criterion is met, then the ring will be stable no matter how $\tau$ fluctuates. The condition is
\begin{equation}
  \zeta < \parens{1 + \frac{3}{128\pi} \frac{\sigma^2}{\Omega^2 R^2}}^{1/3}
  \label{eqn:phase}
\end{equation}
where we have used Eq.~\ref{eqn:sigma} to convert $T$ to velocity dispersion $\sigma$.

Eq~\ref{eqn:phase} forms the solid line boundary of the blue ``moon'' region in Fig.~\ref{fig:phase}b, and the black line in Fig.~\ref{fig:models}. The stability of almost all simulations is accurately predicted by Eq.~\ref{eqn:phase}, demonstrating that the assumptions listed in \S\ref{sec:statistics} may be appropriately applied to planetary rings.

\paragraph{Non-uniform ring phase} While Eq.~\ref{eqn:phase} is the condition for the specific moon-forming mode to be stable, the condition that \textit{all} $k_y=0$ modes are stable is
\begin{equation}
  \zeta < \parens{\frac{\sigma}{6\tau R\Omega}}^{1/3} \brackets{2\pi(1-\tau)}^{-1/6}.
  \label{eqn:a-yellow}
\end{equation}
The Toomre-$Q$ condition also assesses whether all modes are stable, but via the dispersion relation of density fluctuations. That condition has a similar form (Eq.~\ref{eqn:toomre}). When Eq.~\ref{eqn:a-yellow} is violated but Eq.~\ref{eqn:phase} is satisfied, the system occupies the non-uniform ring phase where some modes with $k_x<k_\mathrm{moon}$ are unstable. The yellow region above the dashed line in Fig.~\ref{fig:phase}b depicts this phase using the simulations' $\tau=1/4$.

All simulations in this phase yielded stable rings, indicating that Eq.~\ref{eqn:a-yellow} is not a stability criterion. That is, rings are stable even when there are unstable modes with $k_x < k_\mathrm{moon}$. We suggest that these longer wavelength unstable modes are unable to seed the small over-densities that later collect to form satellites and so do not cause collapse. Instead, they are halted in their exponential growth by some other mechanism. This will cause nonzero particle eccentricities and density correlations in equilibrium. We therefore refer to this state as ``non-uniform'' rings.

One possible halting mechanism is the presence of a minimum in $H_\mathrm{eff}$ at nonzero $c(\bm k)$. For example, if the integrand of Eq.~\ref{eqn:h} were $-2\alpha |c(\bm k)|^2 + \beta|c(\bm k)|^4$, then $c(\bm k)$ would be unstable at zero as it is now, but would grow to the new stable value at $|c(\bm k)|^2 = \alpha/\beta$. This effect could be modeled only if assumption 2 were lifted and higher order terms were included in $H_\mathrm{eff}$. Alternatively $c(\bm k)$ may be halted in its exponential growth by a dynamical, non-equilibrium effect, in which case statistical mechanics is not applicable. Which effect occurs could be probed by measuring $c(\bm k)$ in simulations of rings in this phase.

\paragraph{Triple point} The ``triple point'' where all three phases meet has
\begin{equation}
\sigma_\mathrm{TP} = \parens{\frac{32\sqrt{2\pi}}{9\tau}} \Omega R,\qquad \zeta_\mathrm{TP} = \parens{\frac{16}{27 \tau^2}}^{1/3} \label{eqn:zeta-tp}
\end{equation}
for $\tau \ll 1$, and is marked with a star in Fig.~\ref{fig:phase}b. The location of the triple point helps determine whether the non-uniform phase is accessible. When $\zeta > \zeta_\mathrm{TP}$, the non-uniform phase can occur for the lowest stable velocity dispersions.

\section{Pressure and Azimuthal Asymmetry} \label{sec:pressure}
The techniques of statistical mechanics allows the ring's ``pressure'' $P$ to be computed. One would predict the ring to expand radially if $P>0$ and contract if $P<0$. Pressure can be computed by $P = T\partial \ln Z / \partial A$ where the derivative is computed at constant temperature and particle number $N$ \citep{huang1987statistical}, $A$ is the ring area, and
\begin{equation}
    Z = \sum_\mathcal{S} e^{-H(\mathcal S)/T} \propto A^N \sum_c \sum_h e^{-H_\mathrm{eff}(c, h)/T}
    \label{eqn:partition}
\end{equation}
is the partition function. The $A^N$ factor comes from the $M^{N_b}$ factor discussed in appendix \ref{app:heff}. Note that $Z$ depends on area through the optical depth $\tau$, since increasing ring area while holding the number of particles constant requires optical depth to decrease.

When $\zeta \ll 1$, the orbital energy which encodes particle epicyclic motion dominates the total energy, and gravitational effects are small. In this case, the pressure satisfies
\begin{equation}
  PA \approx \parens{1 + \frac{\pi}{2}} NT.
  \label{eqn:state}
\end{equation}
This formula is the ideal gas law, with an additional $\pi/2$ factor that arises from the somewhat \textit{ad hoc} choice of the large-$k$ cutoff $\Lambda$. Because the pressure is positive, the ring should perpetually expand as found by \citet{lynden1974evolution} and others.

When $\zeta$ is large, self-gravity introduces corrections to Eq.~\ref{eqn:state}. The most severe correction occurs for rings in the non-uniform ring phase. These rings have some modes with effective spring constant $\epsilon_c = 0$, which can achieve arbitrarily high amplitudes without changing the total energy of the ring. To understand the consequences of these modes, we discretize the ring's Fourier modes to a finite number of wavenumbers $\{\bm k\}$, separated by intervals of $\Delta k$. Then the wavenumber integral $\int d^2 \bm k$ in $H_\mathrm{eff}$ becomes a sum $\sum_{\bm k} \Delta k^2$. Applying these substitutions to Eq.~\ref{eqn:partition}, $\ln Z$ is
\begin{equation}
  \begin{aligned}
  \ln Z = N \ln A + \sum_{\bm k}\Bigg[&\ln \sum_{c(\bm k)} \exp\parens{-\frac{\Delta k^2 \epsilon_c(\bm k)}{(2\pi)^2T}|c(\bm k)|^2} \\
  + &\ln \sum_{h(\bm k)} \exp\parens{-\frac{\Delta k^2 \epsilon_h(\bm k)}{(2\pi)^2T}|h(\bm k)|^2}\Bigg]
  \end{aligned}
\end{equation}
where $\sum_{c(\bm k)}$ sums over all possible values of $c(\bm k)$.
For the modes with $\epsilon_c = 0$, the sum over $c(\bm k)$ attains very large values. (The sum technically diverges in the above formula, though the unmodeled higher order terms of $H_\mathrm{eff}$ will cut off the sum over $c$ at a very large value $I$. The value of $I$ is unimportant---this section only requires that it dominates the $\ln Z$ contribution from other modes.) We can therefore neglect all the other modes, simply finding that $\ln Z \approx N_0 \ln I$, where $N_0$ is the number of modes with $\epsilon_c = 0$. 

It follows that $P \propto d N_0 / dA$; that is, the ring's area changes so as to maximize the number of $\epsilon_c = 0$ modes. Inspection of Eq.~\ref{eqn:eps-c} reveals that in strong gravity, $N_0$ increases as $\tau$ increases (i.e., as area decreases). Therefore, the pressure in the non-uniform phase is negative.

If the assumptions listed in \S\ref{sec:statistics} continue to hold for this non-uniform phase, then these rings should shrink over time due to the negative pressure. Occultation observations of Q1R have shown the rings to be azimuthally asymmetric. Most regions of Q1R have low enough optical depth, but small zones have $\tau$ up to 0.4 \citep{pereira2023two}. The ratio of Q1R's $\zeta$ value to the triple point (Eq.~\ref{eqn:zeta-tp}) is
\begin{equation}
  \frac{\zeta}{\zeta_\mathrm{TP}} = \parens{\frac{\tau}{0.18}}^{2/3} \parens{\frac{\rho_m}{0.9\ \mathrm{g}\ \mathrm{cm}^{-3}}}^{1/3},
\end{equation}
indicating that these dense regions with $\tau \gtrsim 0.18$ lie above the triple point and could occupy the non-uniform phase. These large-$\tau$ regions could occur when the optical depth of a small patch of the ring randomly fluctuates upward into the dense non-uniform phase, at which point pressure effects drive the patch to shrink while the surrounding ring expands. This accentuates the initially small density non-uniformity, leading to sporadic patches of strong density. Eventually, shearing could dissipate the patch and halt the shrinking.

This localized pressure effect could act concurrently with other processes known to cause azimuthal asymmetry, such as resonances or net ring eccentricity driven by undiscovered satellites. However, pressure effects should be dissipated over shorter timescales than density variations caused by eccentricity. If the density non-uniformities prove intermittent, pressure effects are a more likely source of the non-uniformities.

However, it is not yet clear whether the assumptions of \S\ref{sec:statistics} do hold. We demonstrated agreement with the simulations of \citet{morgado2023dense} at $\tau = 0.25$, but Fig.~\ref{fig:density} and the discussion of \S\ref{sec:statistics} indicated that the statistical model might not be valid at larger optical depths. Simulations are necessary to determine whether the model can be applied to $\tau \sim 0.4$, non-uniform rings.

\section{Conclusions} \label{sec:conclusion}
Past work has shown that large velocity dispersion can stabilize planetary rings against gravitational collapse, but a correction to the Roche limit that models this effect has not been found. We test several simple models but find that they do not match simulations. We therefore propose a new model which analogizes a ring accreting into moons to a gas condensing into a liquid. Applying statistical mechanics allows the stability criterion to be computed. The result (Eq.~\ref{eqn:phase}) is consistent with simulations of Quaoar's outermost ring. It therefore explains the stability of Quaoar's rings outside the Roche limit, and is the first analytical, first-principles model to do so.

When the ring is outside the Roche limit and is sufficiently optically thick, the model predicts negative pressures due to strong self-gravity, which will shrink dense portions of the ring if the assumptions of the model hold. We argue this effect could exacerbate the azimuthal asymmetry of Quaoar's rings, though ring eccentricity or the influence of satellites could also cause the asymmetry. Further observations of the stability of the asymmetries could address whether the negative pressure effect occurs in real rings.

Resonances can drive ring eccentricity, which can be phase locked by self gravity as first pointed out by \citet{goldreich1979precession,goldreich1979towards} and applied to Chariklo's ring by \citet{pan2016on,melita2020apse}. This statistical theory could also model resonances by including their energy contributions in $H_\mathrm{eff}$. Furthermore, injection of angular momentum can confine rings \citep{goldreich1979towards} and cause them to radially shrink \citep{goldreich1995single} as applied to Chariklo in the absence of self-gravity by \citet{salo2026rings}. This statistical model could be a useful tool to study the influence of perturbing satellites while taking self-gravity into account.

The statistical model assumes the ring system is in equilibrium, contains small density perturbations, and is ergodic. However, non-uniform or optically thick rings are known to violate these assumptions. Indeed, we find that our predictions for the shape of large over-densities in rings near the Roche limit do not match simulations. On the other hand, the agreement between Eq.~\ref{eqn:phase} and simulations suggests that these assumptions may be acceptable when determining the ring's stability. Future applications of this model should carefully check whether the assumptions continue to hold.

\section*{Acknowledgments}
The author is very grateful to Scott Tremaine for helpful discussion and useful suggestions. He also thanks Miroslav Broz, Jack J. Lissauer, Julien de Wit, Laura K. Schaefer, Charles F. Yang, Mason Kamb, Haley R. Stueber, and Roger W. Romani for inspiring conversations, and the anonymous reviewers for their helpful comments.

\bibliographystyle{cas-model2-names}

\bibliography{bib}

\appendix

\section{Energy Penalty for Particle Collisions}
\label{app:heff}
This section determines how to compute Eq.~\ref{eqn:average} while avoiding states in which particles overlap. First we divide the ring into bins indexed by $b$, having position $\bm x_b$, area $A_b$, energy $H_b$, and $N_b$ particles. These bins are sufficiently small that $c(\bm x)$ and $h(\bm x)$ are approximately constant within each bin. The sum over all states in Eq.~\ref{eqn:average} can now be written as a sum over the possible values of $c(\bm x_b)$ and $h(\bm x_b)$ multiplied over all bins
\begin{equation}
  \chevronsi{Q} = \frac{\prod_b \sum_{c(\bm x_b)}\sum_{h(\bm x_b)} X_b Q e^{-H_b/T}}{\prod_b \sum_{c(\bm x_b)} \sum_{h(\bm x_b)} X_b e^{-H_b/T}}
  \label{eqn:avg-intermediate}
\end{equation}
where $X_b$ is the number of ways to distribute the particles within the bin's area without overlapping. To compute $X_b$, suppose that each particle occupies one particle-sized ``slot,'' with area $\pi R^2$ in the orbital plane. The first particle has $M = A_b / (\pi R^2)$ available slots, the next has $M-1$, etc. The number of possible configurations for all $N_b$ particles is
\begin{equation}
  \begin{aligned}
    X_b &= C^M_{N_b}\\
    &\sim M^{N_b}\exp\brackets{-N_b \ln N_b - \parens{M - N_b}\ln \parens{1 - \frac{N_b}{M}}},
  \end{aligned}
\end{equation}
where $C^k_n = k! / [n! (k-n)!]$ is the binomial coefficient. The second line made use of Stirling's approximation $\ln (x!) \sim x \ln x - x$. Defining a density fluctuation $\delta_b$ such that $N_b = M\tau(1 + \delta_b)$, we expand in $\delta_b$. Zeroth order terms are constants and can be neglected for the reasons described in the text surrounding Eq.~\ref{eqn:h}. First order terms average to zero because the bin-averaged value of $\delta_b$ is zero. Third and higher order terms are neglected through assumption 2. So we simply have the second-order term
\begin{equation}
    X_b \sim M^{N_b} \exp\brackets{-\frac{M\tau}{2} \frac{\delta_b^2}{1-\tau}}.
    \label{eqn:ngc}
\end{equation}

When inserting $X_b$ into \ref{eqn:avg-intermediate}, the $M^{N_b}$ factor cancels between the numerator and denominator because $\prod_b M^{N_b} = M^N$ is a constant. We can easily treat the exponential term by defining ``effective energy''
\begin{equation}
  H_\mathrm{eff} = H + T\sum_b \frac{M \tau}{2}\frac{\delta_b^2}{1-\tau}
\end{equation}
and replacing $H$ with $H_\mathrm{eff}$ in the Boltzmann factor (e.g.~Eq.~\ref{eqn:average2}).

Now we take the limit as the bins become small, so $\sum_b M = \sum_b A_b / (\pi R^2)$ becomes an integral $\int d^2\bm x / (\pi R^2)$. In this limit, the effective energy is
\begin{equation}
  H_\mathrm{eff} = H + \frac{T\tau}{2\pi R^2}\frac{1}{1-\tau}\int d^2\bm x \abs{\frac{\delta n(\bm x)}{n_0}}^2,
  \label{eqn:h-s}
\end{equation}
where $\delta n(\bm x)$ is the ring's local surface number density perturbation, and the average number density is $n_0 = \tau / (\pi R^2)$.

To compute $\delta n(\bm x)$, we note the definition
\begin{equation}
  \delta n(\bm k) = n_0\int d^2\bm x\, \brackets{e^{-i\bm k \cdot (\bm x + \bm r(\bm x))} - e^{-i\bm k \cdot \bm x}}.
  \label{eqn:n-def}
\end{equation}
If displacements $\bm r$ are small enough, one may expand in $\bm k \cdot \bm r$ to achieve $\delta n(\bm k) = -i n_0\bm k \cdot \bm r(\bm k)$, but this expansion breaks down for $\bm k \cdot \bm r \sim 1$ because $e^{-i\bm k \cdot \bm r}$ is oscillatory and cancels with itself over the integral. To compute the result in this large-$k$ limit, note that the power spectrum $|\delta n(\bm k)/n_0|^2 = \int d^2\bm x\, d^2 \bm y e^{-i\bm k\cdot (\bm x-\bm y)}e^{-i\bm k\cdot (\bm r(\bm x)-\bm r(\bm y))}$ receives coherent contributions only when the second exponent is zero, i.e.~$\bm x=\bm y$. We discretize this integral into chunks of area $\Delta x^2$ by replacing $\int d^2 \bm x$ with $\sum_x \Delta x^2$ where $\Delta x^2 = 1/n_0$,  since $1/n_0$ is the average ring area per particle. Summing only over the terms where $\bm x = \bm y$, we reach $|\delta n(\bm k)/n_0|^2 = A \pi R^2 / \tau$  in the large-$k$ limit, where $A$ is the ring's area.

A fully accurate model would quantify the transition between the small- and large-$k$ regimes, but as an approximate approach we set $|\delta n(\bm k)|^2$ to transition abruptly between the two. Using Eq.~\ref{eqn:displacement} to convert $\bm r$ to the $c$-field,
\begin{equation}
  \abs{\frac{\delta n(\bm k)}{n_0}}^2 = \min\brackets{\frac{a^2|c(\bm k)|^2}{2}(k_x^2 + 4k_y^2), A\frac{\pi R^2}{\tau} }.
  \label{eqn:n}
\end{equation}

\section{Calculation of Energy}
\label{app:h}
The energy $H$ is a sum of orbital energy $H_O$, which incorporates the role of epicyclic motion, and gravitational potential energy $H_U$, which incorporates self-gravity.

The ring's total orbital energy is $H_O = -\sum_jGMm/(2a_j)$, where $j$ indexes over particles and $a_j$ is a particle's semi-major axis. Recall that $a_j = |\bm L_j|^2 / [GM(1 - e_j^2)]$, where $\bm L_j$ is the orbital angular momentum. Since the guiding center has orbital angular momentum equal to $L_{j,z} = |\bm L_j|\cos\iota_j$ by definition, the guiding center has semi-major axis $|\bm L_j|^2\cos^2 \iota_j  / (GM)$. This semi-major axis is also equal to $a + s_{j,x}$ for guiding center position $\bm s_j$. Eliminating $\bm L_j$, one obtains
\begin{equation}
  H_O = \frac{m\Omega^2a^2}{2} \sum_j \brackets{|c_j|^2 +|h_j|^2-\parens{1 - \frac{s_{j,x}}{a} + \frac{s_{j,x}^2}{a^2}}}
  \label{eqn:h-orb-particle}
\end{equation}
to second order in $c_j$, $h_j$, and $\bm s_j$, where $\Omega^2 = GM/a^3$. We have used the fact that $|c_j|^2 = e_j^2$ and $|h_j|^2 = \iota_j^2$.

Recall that we chose not to model the distribution of guiding centers as statistical (\S\ref{sec:dof}). The third term depends only on the guiding center locations, and is therefore not relevant to determining the statistical distribution of the $c$ and $h$ fields we focus on. Furthermore, this term is effectively constant. As particles interact, the values of $s_{j,x}$ (and the other orbital elements) will change for each particle. However, $H_O$ depends only on $\sum_j s_{j,x}$ and $\sum_j s_{j,x}^2$ which evolve very slowly. In particular, $\sum_j s_{j,x}=0$ is constant since the origin was chosen to lie in the center of the ring, and $\sum_j s_{j,x}^2$ is proportional to the squared width of the ring, which expands over time scales far longer than the orbital period \citep{lynden1974evolution}. As explained in the text surrounding Eq.~\ref{eqn:h}, terms in $H$ which do not depend on $c$ or $h$ and are constant in time can be neglected. The orbital energy contributed by the remaining terms is
\begin{equation}
  \begin{aligned}
    H_O &= \frac{m\Omega^2a^2\tau}{2\pi R^2}\int d^2\bm x\, \brackets{|c(\bm x)|^2 + |h(\bm x)|^2}\\
    &= \frac{m\Omega^2a^2\tau}{2\pi R^2}\int \frac{d^2\bm k}{(2\pi)^2}\, \brackets{|c(\bm k)|^2 + |h(\bm k)|^2}.
    \label{eqn:h-o}
  \end{aligned}
\end{equation}

The ring's gravitational energy is obtained by integrating the potential energy between all particle pairs:
\begin{equation}
  \begin{aligned}
  H_U = -\frac{Gm^2}{2}\int d^2\bm x\, d^2\bm y\, &\frac{n(\bm x) n(\bm y)}{\sqrt{|\bm x - \bm y|^2 + |r_z(\bm x) - r_z(\bm y)|^2}}\\
  \approx -\frac{Gm^2}{2}\int d^2\bm x\, d^2\bm y\, &\bigg[\frac{\delta n(\bm x) \delta n(\bm y)}{|\bm x - \bm y|} \\
  &- n_0^2\frac{(r_z(\bm x) - r_z(\bm y))^2}{2|\bm x - \bm y|^3}\bigg] + U_0
  \label{eqn:potential-space}
  \end{aligned}
\end{equation}
where $\bm r$ was given in Eq.~\ref{eqn:displacement}. The second equality results from a leading order Taylor expansion. Our decision not to include higher order terms invokes assumption 2, since such terms are at least second-order in density. The last term in Eq.~\ref{eqn:potential-space}, $U_0$, represents the self-gravitational energy of a uniform ring. This constant does not depend on $c$ and $h$ and can therefore be neglected like the first term of Eq.~\ref{eqn:h-orb-particle}.

Gravitational interactions lead to terms in $H_U$ that mix fluctuations at different locations, which would complicate the analysis later on. We avoid this complexity by switching to Fourier space, in which the modes work out to be uncorrelated to leading order. The Fourier transform of $1/|\bm x - \bm y|$ is $2\pi/|\bm k|$, and that of $1/|\bm x - \bm y|^3$ is $-2\pi |\bm k|$. Applying these identities and the convolution theorem, $H_U$ is in Fourier space
\begin{equation}
  H_U = \frac{12m\Omega^2\tau^2\zeta^3}{\pi R}  \int \frac{d^2\bm k}{(2\pi)^2}\brackets{|\bm k| \abs{r_z(\bm k)}^2-\frac{1}{|\bm k|}\abs{\frac{\delta n(\bm k)}{n_0}}^2}.
  \label{eqn:h-u}
\end{equation}
Using Eq.~\ref{eqn:displacement} to express $\bm r$ in terms of $c$ and $h$, the total energy $H = H_O + H_U$ can be computed using Eqs.~\ref{eqn:h-o} and \ref{eqn:h-u} in terms of $|c(\bm k)|^2$, $|h(\bm k)|^2$, and the density power spectrum $|\delta n(\bm k) / n_0|^2$. The terms proportional to $|c(\bm k)|^2$ and $|h(\bm k)|^2$ are factored into $\epsilon_c(\bm k)$ and $\epsilon_h(\bm k)$  given in the main text at Eqs.~\ref{eqn:eps-c} and \ref{eqn:eps-h}. Eq.~\ref{eqn:n} also gives $|\delta n(\bm k) / n_0|^2$ in terms of $|c(\bm k)|^2$, but the minimum function therein must be properly treated. When the first argument of the minimum is smaller, $|\delta n(\bm k) / n_0|^2 \propto |c(\bm k)|^2$ and should therefore be included in $\epsilon_c(\bm k)$. When the second argument is smaller, $|\delta n(\bm k) / n_0|^2$ is independent of $c$ and should be neglected as constant. We use the symbol $\mathbb{1}(\bm k)$ to distinguish between these cases in Eq.~\ref{eqn:eps-c}; $\mathbb{1}(\bm k)=1$ when the first argument of Eq.~\ref{eqn:n} is smaller and zero otherwise.

In equilibrium, one can replace $|c(\bm k)|^2$ in Eq.~\ref{eqn:n} with $\chevronsi{|c(\bm k)|^2}$ (Eq.~\ref{eqn:exp}) to evaluate $\mathbb{1}(\bm k)$. The result is $\mathbb{1}(\bm k)=0$ when $k_x^2 + 4k_y^2 > m\Omega^2 / T$, as stated in the main text. However the stability criterion discussion of \S\ref{sec:roche} explicitly considers whether unstable \textit{non-equilibrium} configurations exist. For any wavenumber, one can construct a non-equilibrium configuration where the first argument of the minimum should be taken.\footnote{For example, consider the case where $c(\bm k)=0$ for all modes except those in a small region of area $\Delta k^2$ in Fourier space, centered on $\bm k_0$. Then $r(\bm x) = \int \frac{d^2 \bm k}{(2\pi)^2} c(\bm k) e^{i\bm k \cdot \bm x} = \frac{\Delta k^2}{(2\pi)^2} c(\bm k_0) e^{i\bm k_0 \cdot \bm x}$. Because $\Delta k^2$ is small, $\bm r(\bm x)$ remains small no matter the size of $c(\bm k_0)$. One can therefore Taylor expand the $e^{i\bm k \cdot \bm r(\bm x)}$ term of Eq.~\ref{eqn:n-def}, which gives the first argument of the minimum in Eq.~\ref{eqn:n}.} Therefore, we always set $\mathbb{1}(\bm k)=1$ in that section.

\end{document}